# Calcite single crystals as hosts for atomic-scale entrapment and slow release of drugs

*Giulia Magnabosco[†], Matteo Di Giosia[†], Iryna Polishchuk[‡], Eva Weber[‡], Simona Fermani[†], Andrea Bottoni[†], Francesco Zerbetto[†], Boaz Pokroy[‡,\*], Stefania Rapino[†,∥,\*], Giuseppe Falini[†,\*], Matteo Calvaresi[†,\*]*

[†] Dipartimento di Chimica "Giacomo Ciamician", Alma Mater Studiorum Università di Bologna, via Selmi 2, 40126, Bologna, Italy

[‡] Department of Material Sciences and Engineering and the Russell Berrie Nanotechnology Institute, Technion – Israel Institute of Technology, 32000 Haifa, Israel

[∥] Istituto Europeo di Oncologia, via Ripamonti 435 Milano, Italy

ABSTRACT. This study presents a complete structural and biological characterization of Doxorubicin/$CaCO_3$ single crystals as a pH responsive drug carrier. Using a biomimetic approach, it was demonstrated that calcite single crystals are able, during their growth in presence of doxorubicin, to entrap drug molecules inside their lattice, along specific crystallographic directions. High-resolution synchrotron powder diffraction measurements allowed the determination of the lattice distortion and microstructural parameters. Confocal microscopy confirmed that doxorubicin is uniformly embedded in the crystal and that the drug is not only adsorbed on the crystal surface. A slow release of DOX is obtained that occurs preferentially in proximity of the crystals, targeting cancer cells.

Nano- and micro-particles hold great promise for controlled and targeted drug release and delivery.[1-3] An ideal drug carrier should not exert harmful effects on normal cells. It should also satisfy requirements of stability, *in vivo* biocompatibility, and ability of targeted on-demand release.[1-3] Inorganic nanomaterials may fulfill most of these requirements. Due to the simplicity of synthesis and modification, it is possible to control the particle size, shape and surface functionalization.[3] They are usually made of durable and robust materials, which allow encapsulation and protection of sufficient amounts of cargos, preventing pre-leakage and damage to normal cells.[4] Among inorganic carriers, calcium carbonate crystals represent an ideal platform as smart carriers for drugs due to their capability to adsorb,[5-16] and more importantly to entrap, drugs.[17] Organic molecules are commonly found incorporated inside biogenic calcium carbonate ($CaCO_3$) crystals.[18,19] Nature in several organisms uses minerals to store vital ions and structural and functional molecules and macromolecules.[20,21] In this view the use of $CaCO_3$ as cargo-carrier represents an application of bio-mineralization inspired principles.[22-24]

$CaCO_3$ forms common biominerals that are intimately associated to biological fluids, which, in turn, makes them highly biocompatible. In the body, it degrades to $Ca^{2+}$ and $CO_3^{2-}$, which are products that do not pose problems of toxicity.[25, 26]

The preparation of $CaCO_3$ crystals is a simple, low cost and organic solvent-free process (low-level exposure to residual toxic organic solvents may lead to lasting toxic effects). Under highly optimized conditions, crystal formation can control the size of $CaCO_3$ particles from the micro- to the nano-meter range, with relatively narrow size distributions.[27] The pH sensitive $CaCO_3$ solubility can release entrapped molecules only when the dissolution of the crystals occurs.[28] Drug/$CaCO_3$ hybrid crystals allows zero-leakage of drugs at the physiological pH of 7.4, and release drugs in acidic conditions. CaCO$_3$ carriers are particularly suitable for selective release of drugs in tissues that are more acidic than normal physiological pH (tumors, inflamed tissues).[29,30]

In general, development of an effective drug delivery system requires understanding of the chemical and physical properties that affect *i)* the interaction of the drug with the micro- nano- particles, and *ii)* the interaction of the micro- nano- carries with the biological environment. Often, the structural characterization of the interaction between drug and carrier is missing. Dopant molecules can be either located between individual crystallites of polycrystalline materials or entrapped inside single crystals where they can interact with specific crystallographic patterns.

In the present work, doxorubicin (DOX), an anthracycline drug widely used in chemotherapy,[31] is used as model molecule to study the entrapment in calcite crystals and its release. The aim is to determine at what concentrations DOX is incorporated inside CaCO$_3$ single crystals, examine the effects of such incorporation on the hosting crystalline lattice and understand the mechanism of incorporation. The biological activity of CaCO$_3$/DOX systems is also assayed.

RESULTS AND DISCUSSIONS

Calcium carbonate/DOX hybrid crystal precipitation was conducted at room temperature by controlled diffusion of CO$_2$ and NH$_3$ vapors into 10 mM calcium chloride solutions containing different concentrations of DOX. The precipitation process was stopped after 4 days (see Methods for details).

**Effect of DOX doping on the morphology of calcite crystals.** In the absence of DOX, only rhombohedral single crystals of calcite are precipitated. In these crystals, characterized by an average size of 50 μm, only the typical {10.4} faces are observable. The presence of DOX influenced the crystallization process as a function of its initial concentration in solution. At low concentrations, 5·10$^{-4}$ mM and 5·10$^{-3}$ mM, DOX did not affect the precipitation of calcite. At a concentration of 5·10$^{-2}$ mM the crystals became hoppered, showing holes on the {10.4}rhombohedral faces. Increasing the concentration of DOX to 5·10$^{-1}$ mM and 5 mM results in the observation of the co-presence of spherulites with aggregated rhombohedral crystals and in the strong inhibition of the precipitation that is usually associated with the deposition of few aggregates and submicron sized particles. Calcite was the only crystalline phase detected by X-ray powder diffraction in the control experiment and in the experiments in which 5·10$^{-4}$ mM, 5·10$^{-3}$ mM or 5·10$^{-2}$ mM DOX was used. When a higher concentration of DOX was used (5·10$^{-1}$ mM and 5 mM), small amounts of vaterite co-precipitated with calcite (Fig. 1). This preliminary investigation indicated that 5·10$^{-2}$ mM DOX was the optimal initial concentration for the aims of this research.

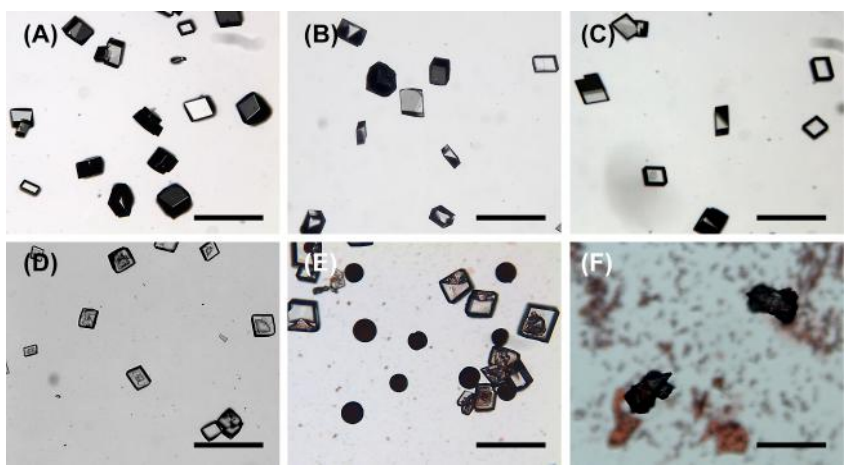

**Figure 1** Optical microscope pictures of DOX/calcite hybrid crystals precipitated in the presence of different initial concentrations of DOX. (A) no DOX, (B) 5·10$^{-4}$ mM, (C) 5·10$^{-3}$ mM, (D) 5·10$^{-2}$ mM, (E) 5·10$^{-1}$ mM and (F) 5 mM. Scale bar 100 μm.

The textural features of the DOX/calcite hybrid crystals were further investigated by scanning electron microscopy, as illustrated in Figure 2.

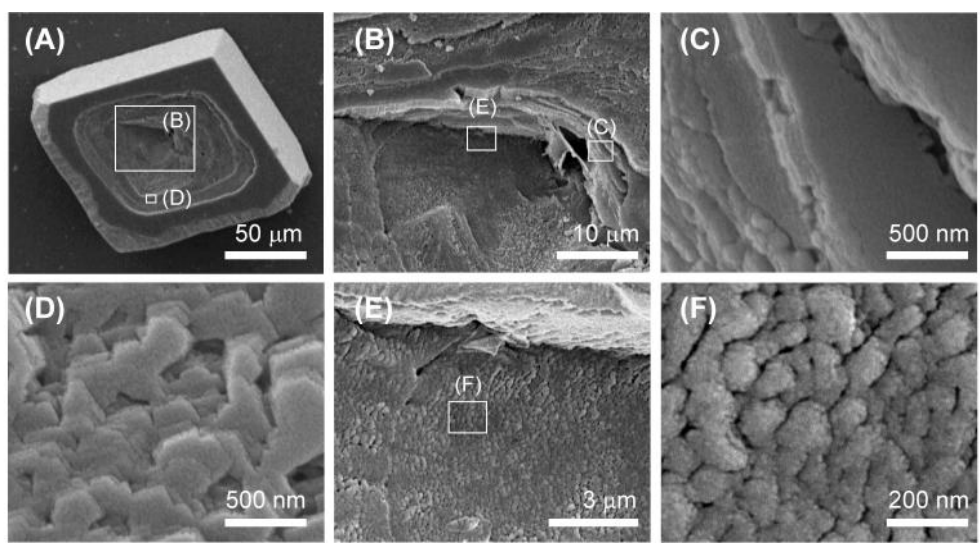

**Figure 2** Scanning electron microscope pictures of DOX/calcite hybrid crystals precipitated in the presence of $5 \cdot 10^{-2}$ mM DOX. Textural details of different region of the crystal in (A) are illustrated. The rectangle indicates the region of magnification and the subscript the corresponding picture.

Each crystal was characterized by the presence of a rounded cavity in the center of one of the {10.4} face. The diameter of this cavity increased from inside to outside and changed among crystals. The wall of the hole was stepped, with each step formed by a flat (10.4) face and a rough (*hk.l*) face. The thickness of the steps decreased from about 500 nm to less than 200 nm moving toward the core of the crystal. The surface of the (*hk.l*) face showed the presence of packed spheroid nanoparticles, of about 100 nm, on which particles of few nanometers were observed. The formation of hoppers was explained by growth spiraling outwards from a screw dislocation[32] (thermodynamic control) and by limited diffusion of constituent ions to the growing crystal face[33] (kinetic control). In this case, formation of additional crystalline faces on the hole walls supports a mechanism of thermodynamic control of growth spiraling due to the screw dislocations induced by the DOX – calcite interaction. This aspect was further investigated by X-ray diffraction analysis.

**Effect of DOX doping on the lattice structure of calcite crystals.** To examine the influence of DOX on the crystal structure of calcite hybrid crystals in the presence of the optimal concentration $5 \cdot 10^{-2}$ mM of DOX, high-resolution synchrotron powder diffraction (HRXRD) measurements were carried out on the 11-BM beamline (Argonne National Laboratory, Argonne, USA). These measurements allowed us, by determination of lattice distortions, to ascertain whether DOX indeed is incorporated into the calcite lattice. We have already demonstrated that when organic molecules are incorporated into an inorganic crystalline host they induce lattice distortions[34-37,] and lead to unique microstructures.[38,39] This has been shown both in biogenic crystals as well as in bio-inspired calcite[40] and ZnO.[41,42] The procedure of measurement has been described extensively elsewhere. In short, we applied Rietveld analysis and line profile analysis on the full diffraction patterns.

The measurements were performed on the control calcite sample, DOX/calcite hybrid sample and on the DOX/calcite hybrid sample after a mild thermal annealing at 250° C for 2h. Analysis of the diffraction patterns indicates a single calcite phase in all measured samples (see Figure 3 as an example).

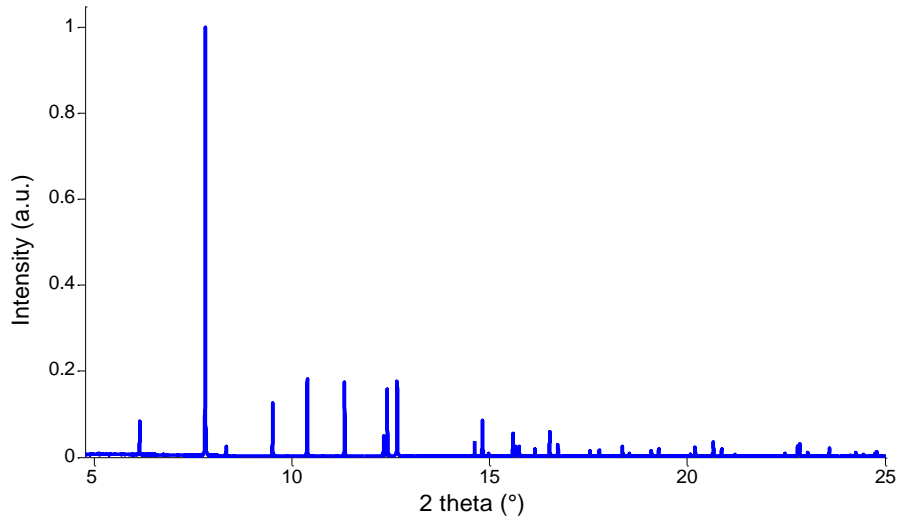

**Figure 3.** HRXRD diffraction profile of the DOX/calcite hybrid crystals precipitated in the presence of $5 \cdot 10^{-2}$ mM DOX.

Observing the (104) diffraction peak of DOX/calcite hybrid crystals as compared to the control sample demonstrates a clear peak shift towards a smaller Bragg angle, which indicates lattice expansion (see Figure 4). This lattice distortion is due to the incorporation of DOX inside the calcite crystal. In Figure 4 the (104) diffraction peak of the hybrid crystal is split. The splitting is due to the fact that the majority of the crystals well incorporate DOX, while the minority of the crystals do not. After the mild annealing treatment the diffraction peak shifts back to that of the control position due to lattice distortion relaxation as the organics are burnt out. It can also be seen that the post-annealing peak becomes considerably wider, the FWHM doubles (from 0.006° to 0.0126°) and is symmetrical. This latter phenomenon was also observed in previous cases such as for biogenic crystals[38] and for crystals that contained amino acids incorporation.[40,42] The cause of the broadening after annealing has been discussed previously, but in short, stems from the formation of new interfaces as the organics are burnt out.

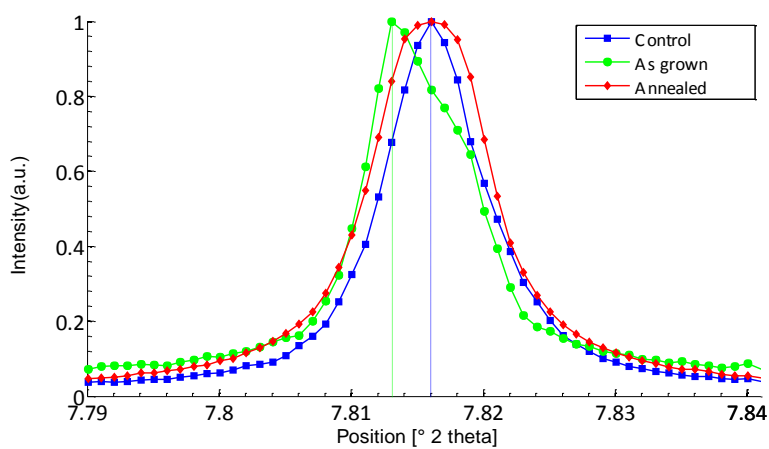

**Figure 4.** The (104) calcite diffraction peak for control calcite (blue), DOX/calcite hybrid crystals (green) and after thermal annealing (red).

Rietveld structure refinement analysis was applied to all diffraction spectra utilizing the GSAS software[43] and the EXPGUI interface.[44] This analysis provided lattice distortion determination by comparing the lattice parameters to that of the control sample. The summary of the lattice distortions and unit cell volumes are listed in Table 1. The strain along the *a* and *c* axes are almost equivalent (~$8\cdot10^{-4}$) and of positive sign (i.e., expansion) (see Figure 5). The magnitude of the distortions is similar to what we find both in biogenic crystals and in other biomimetic crystals. After the annealing process, the lattice distortions drop practically to zero within our statistical error. This is due to the complete destruction of the organic molecules within the crystalline lattice. The volume change due to the incorporation is 0.24 %, which might imply that the level of DOX incorporation is about 0.24% in volume.

**Table 1.** Quantitative data of lattice parameters, lattice distortions, unit cell volume and goodness of fit parameters for the Rietveld refinement fit ($\chi^2$).

| DOX concentration in solution (mM) | a, b parameters (Å) | strain a-axis | c parameter (Å) | strain c-axis | unit cell volume (Å$^3$) | $\chi^2$ |
|---|---|---|---|---|---|---|
| - | 4.990380 | - | 17.063950 | - | 368.03 | 4.334 |
| 5·10$^{-2}$ | 4.994405 | 8.066E-04 | 17.077791 | 8.111E-04 | 368.92 | 2.303 |
| 5·10$^{-2}$ Annealed | 4.990319 | -1.222E-05 | 17.064249 | 1.752E-05 | 368.02 | 3.754 |

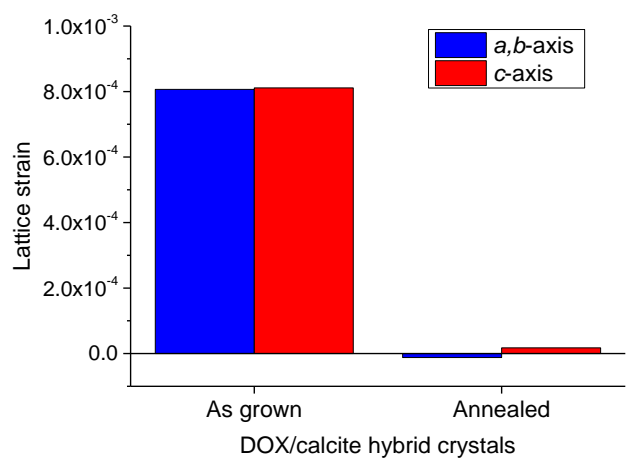

**Figure 5.** Lattice distortions before and after thermal annealing for DOX/calcite hybrids as compared to calcite control.

We further performed line profile analysis on the diffraction spectra. This allowed extraction of microstructural parameters (coherence length and micro-strain fluctuations). Single diffraction peaks were fitted to a Voigt function, which enabled independent evaluation of the contributions of the Lorentzian and Gaussian types, which correlate to the coherence length (crystalline size) and micro-strain fluctuations respectively. The profile fitting was performed using the Gnuplot 4.7 interface[45] over the most intense calcite (104) peak (Figure 6). The results revealed noticeable reduction in crystallite size (threefold) upon annealing accompanied by an increase in the averaged micro-strain fluctuations. This latter finding coincides with that observed for biogenic and other biomimetic crystals in which intra-crystalline organic molecules exist.

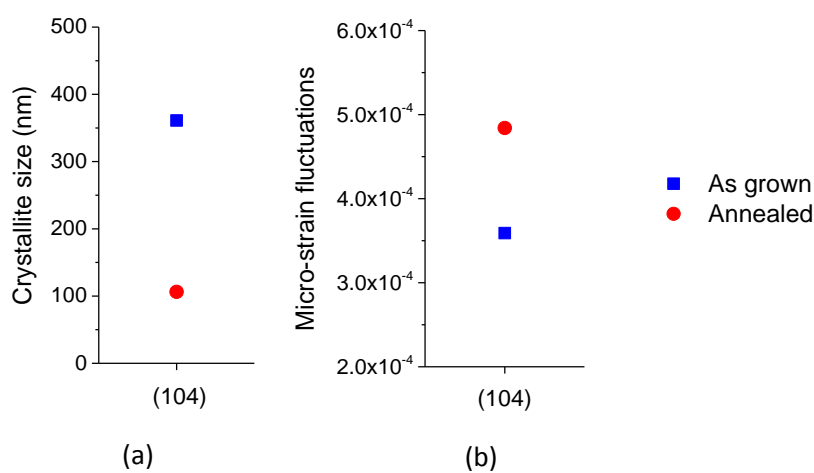

**Figure 6.** Crystallite size (nm) (a) and micro-strain fluctuations (b) before (blue square) and after (red circle) thermal annealing at 250 °C for 120 min for the DOX/calcite hybrid crystals along the (104) crystallographic plane.

**Quantification and distribution of DOX in the DOX/calcite hybrid crystals.**
An evaluation of the total amount of DOX adsorbed in the calcite crystal was carried out by combining UV-vis spectroscopy, for the determination of DOX, and flame atomic absorption spectroscopy, for the determination of $Ca^{2+}$. A loading of $0.3 \pm 0.1$ % (*w/w*) was calculated for the DOX/calcite hybrid crystals.

The spatial distribution of DOX in calcite crystals was also evaluated by assessing the DOX fluorescence by confocal laser scanning microscopy. Figure 7 shows the fluorescence images obtained by a z-stacking of DOX containing crystals. All longitudinal sections of the crystal display a homogeneous fluorescence intensity indicating that DOX was uniformly embedded in the crystal and that the drug is not only adsorbed on the crystal surface. No luminescence could be detected from the reference crystal grown in the absence of DOX

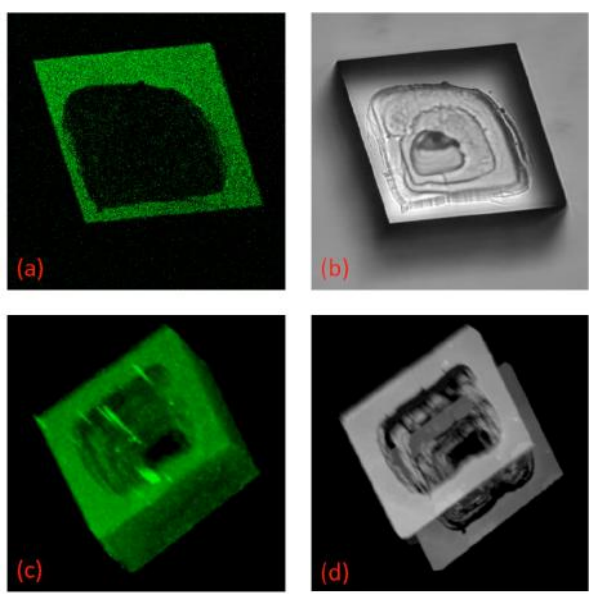

**Figure 7.** Single xy plane (a,b) and 3D reconstruction (c,d) of a z-stacking of a DOX/calcite crystal in fluorescence (a, c) and reflection mode (b, d). The photo-detector was set up in the wavelength range of the DOX emission and of the DOX excitation.

**Targeted DOX release.** As expected, the drug carrier was pH-sensitive (Fig. 8). The DOX release kinetics from DOX/calcite crystals was measured by UV-Vis spectrophotometry in citrate buffer at pH 5.6. The release of DOX was still active after 72 hrs. The same measurements performed at pH 7.4 in PBS did not show any detectable release of DOX.

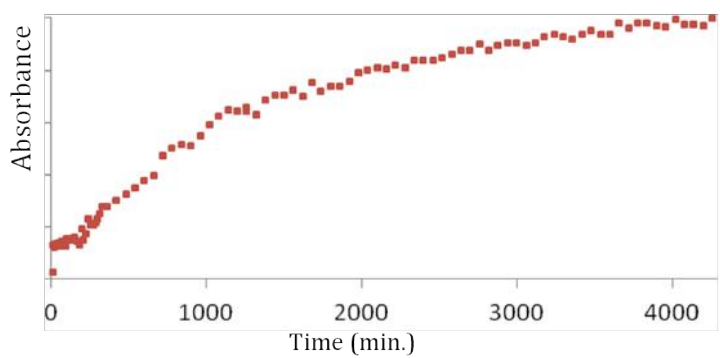

**Figure 8.** Kinetics of DOX release from DOX/calcite hybrid crystals in citrate buffer at pH 5.6.

The drug release from the hybrid crystals is controlled by the dissolution rate of the $CaCO_3$ host crystals.[28]

***In vitro* test of the DOX/calcite hybrid crystals.** The pharmacological activity of DOX/calcite hybrid crystals was tested *in vitro* on cell cultures. The *in vitro* cancer model used is based on the activation of the Ras oncogenic pathway. Ras oncogene is frequently mutated in human cancers and the activation of the Ras signaling pathway provides a good experimental model to study the molecular mechanisms of cancer transformation. Human non-tumorigenic breast epithelial cells (MCF10A) were subjected to retroviral (pBabe vector) infection to express the oncogenic form of Ras, Ha-RasVal12 (RasV12), and the infected/RasV12 expressing cells were selected by puromycin. RasV12 was efficiently expressed in infected cells, leading to the activation of Ras signaling pathway including the downstream effector Erk kinase. Phenotypically, RasV12 expression resulted in altered cell morphology, filopodia-like structure and loss of contact inhibition, soft agar colony formation as well as *in vivo* tumorigenicity.[46]

To tested the toxic effect of DOX/calcite hybrid crystals on transformed MCF10A cells, we cultured the cells in the medium in the presence of DOX/calcite hybrid crystals or calcite crystals as control. Cell growing was followed by cell counting. The contrast phase optical microscopy images at 24h and 72h from plating are presented in Figure 9.

Pure calcite crystals revealed no toxic effect on MCF10A transformed cells. MCF10A cells in the presence of these crystals followed the same cell growth curve of the control cells. In contrast the DOX/calcite hybrid crystals demonstrated heavy toxic effects for the cells (Fig. 9) and after 72 hours the mortality reached 100%. This result suggests that the activity of DOX is retained in the crystal and that the DOX/calcite hybrid crystals are able to release the drug in proper concentrations.

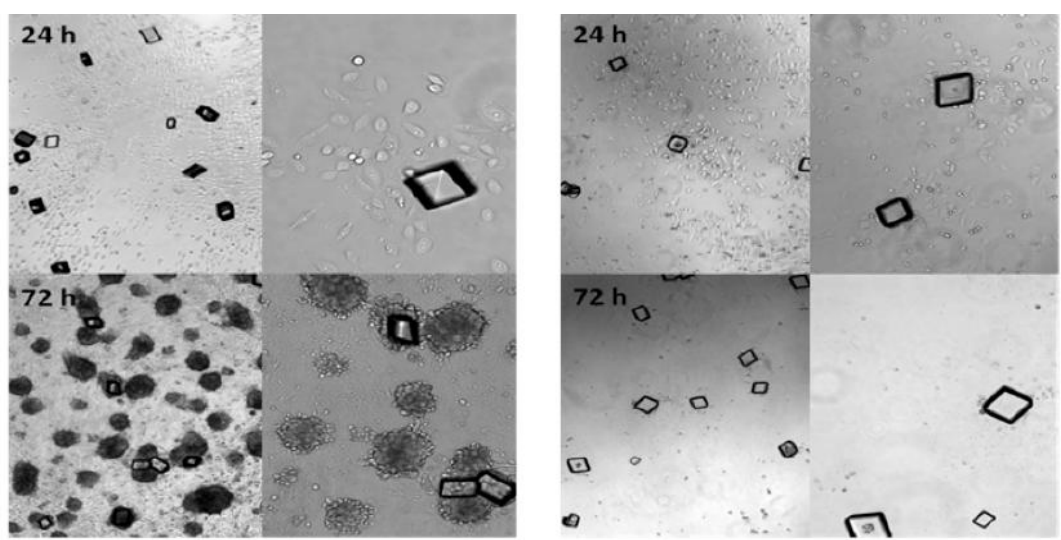

**Figure 9.** MCF10A RasV12 cells cultured with calcite crystals (left) and DOX/calcite hybrid crystals (right). Acquisitions were carried out at 24 and 72 hours.

The DOX uptake by MCF10A RasV12 cells from the DOX/calcite hybrid crystals was assessed by following DOX fluorescence signal, using confocal laser fluorescence microscopy (Figure 10). After 6 hrs of cell culturing in the presence of the crystals, the RasV12 MCF10A cell fluorescence images clearly indicated that the DOX molecules were inside the cells. The DOX signal was clear on the dissolving surfaces of the DOX/calcite crystals (see arrows in Fig. 10). The molecular distribution in the cells is in accord with previous results for the same treatment time.[47]

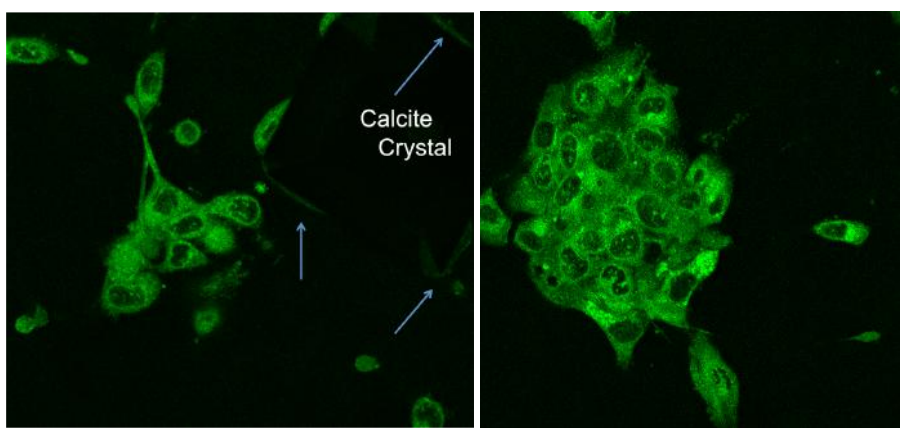

**Figure 10.** Uptake of DOX by cancer cells. The light blue arrows highlight the DOX release by the DOX/Calcite crystal. The arrows indicate the periphery of the calcite crystals.

CONCLUSION

In conclusion, this study presents a complete structural and biological characterization of DOX/CaCO$_3$ crystals as a system to specifically target drugs to particular cells or tissues. The pH sensitive CaCO$_3$ solubility can release entrapped molecules only when the dissolution of the crystals occurs and allows zero-leakage of drugs at the physiological pH. The main results can be summarized in the following points: i) calcite is able to host DOX molecules efficiently; ii) the entrapment occurs along specific crystallographic directions; iii) the release of DOX is controlled by pH and occurs preferentially in proximity of the surface of cancer cells; iv) the released drug molecules are uptake by the cancer cells, killing them.

METHODS

**CaCO$_3$ crystallization experiments.** A 30x30x50 cm$^3$ crystallization chamber was used. Two 25 mL beakers half-full of (NH$_4$)$_2$CO$_3$ (Carlo Erba) and two Petri dishes (d = 8 cm) full of anhydrous CaCl$_2$ (Fluka) were placed inside the chamber. Microplates for cellular culture (Microplate 24 well with Lid, IWAKI) containing a round glass cover slip in each well were used. Into each well, 750 μL of 10 mM CaCl$_2$ solutions were poured. In the experiment with DEOX the required amont of DEOX was dissolved in the 10 mM CaCl$_2$ solution. The micro-plate was covered with aluminum foil and a hole was made over every well. After 4 days the crystals were washed three times with milli-Q water (resistivity 18.2 MΩ cm at 25 °C; filtered through a 0.22 μm membrane) and then analyzed. All the experiments were conducted at room temperature. The crystallization trials of CaCO$_3$ in the different conditions were replicated three times.

**Microscopic observations.** The optical microscope observations of CaCO$_3$ precipitates were made with a Leica microscope equipped with a digital camera. The SEM observations were conducted in a scansion electronic microscope using a Phenom$^{TM}$ microscope (FEI) for uncoated samples and a Hitachi FEG 6400 microscope for samples after coating with gold.

**Atomic absorption spectroscopy.** Atomic absorption measurements of calcium and magnesium were carried out with Perkin Elmer AAnalyst 100 flame and graphite furnace (HGA 800) spectrometer equipped with a Zeeman effect background corrector, and an automatic data processor. A 20-μl volume sample solution obtained by precipitated dissolution in 0.1 M HNO$_3$, was injected by an auto sampler. A multi element hollow cathode lamp of analytes was used as radiation source. Three measurements were carried out for each sample.

**X-ray diffraction analysis.** X-ray diffraction analysis was performed utilizing high resolution synchrotron powder diffraction instrument: 11-BM beamline at Argonne's Advanced Photon Source (Argonne National Laboratory, Argonne, USA). Collection temperature 295.0 K, calibrated wavelength 0.413842 Å. Powders were packed and sealed into polyimide tubes and placed in the beam. Data collected from 12 crystal two-axis analyzer detector.

**Confocal laser scanning microscopy.** Confocal laser scanning microscopy analysis was carried out on cells cultured onto glass coverslips, fixed with 3% paraformaldehyde (Sigma) in PBS and mounted with an antifade glycerol-based medium. Samples were observed with a LEICA TCS SP2 confocal laser scanning microscope (Leica Instruments) without any further modification, the fluorescence of DOX molecule was used to obtain the fluorescence images.

**Measurements of the kinetics of drug release.** The kinetic of DOX release was studied by UV-Vis spectroscopy (Perkin-Elmer Lambda 45). The crystals dissolution was conducted in a 0.5 M citrate buffer solution at pH 5.6 following the absorption intensity of DOX molecules at 499 nm.

**Cell culture.** MCF10A cells (ATCC: crl-10317) were cultured in (1:1) Dulbecco's Modified Eagle's Medium (DMEM) / Nutrient Mixture F-12 Ham (Gibco-Life Technologies Corporation) supplemented with 5% horse serum, 20 ng/ml epidermal growth factor (EGF), 50 ng/ml cholera toxin, 500 ng/ml hydrocortisone and 0.01 mg/ml insulin (Gibco-Life Technologies Corporation)[2]. Cells were infected using a puromycin-resistant retroviral construct containing an oncogenic form of Ras (pBabe-RasV12) or using the empty vector (pBabe) as described in (REF). Forty-eight hours post infection cells were selected using 2 μM puromycin for 4 days.

*Acknowledgment.* B.P. acknowledges funding from the European Research Council under the European Union's Seventh Framework Program (FP/2007–2013)/ERC Grant Agreement n° [336077]. Use of the Advanced Photon Source at Argonne National Laboratory (11-BM beamline) was supported by the U. S. Department of Energy, Office of Science, Office of Basic Energy Sciences, under Contract No. DE-AC02-06CH11357.

We thank Anastasia Brif for her contribution in a part of the data analysis.

**Corresponding Author**

* BP: bpokroy@tx.technion.ac.il; SR: stefania.rapino3@unibo.it;

GF: giuseppe.falini@unibo.it; MC: matteo.calvaresi3@unibo.it

**Author Contributions**

The manuscript was written through contributions of all authors.

**Graphical Table of Contents**

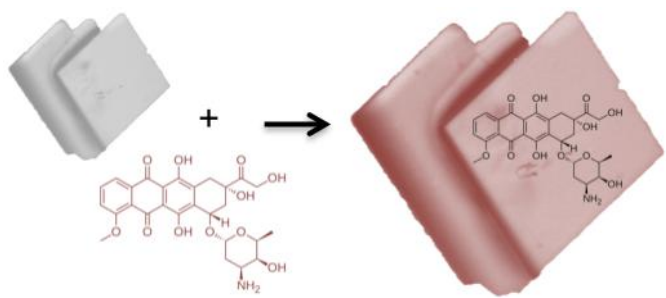